\documentclass[aps,pra,twocolumn,superscriptaddress,showpacs,showkeys,amsmath,amssymb]{revtex4}

\usepackage{amsfonts}
\usepackage{amssymb,amsmath}
\usepackage{mathrsfs}
\usepackage{latexsym}
\usepackage{amsmath}
\usepackage[cp1251]{inputenc}
\usepackage{graphicx}
\usepackage{dcolumn}
\usepackage{bm}
\usepackage{color}

\RequirePackage{ifthen}
\RequirePackage[pdfstartview=FitH]{hyperref}
\begin{document}

    \title{Trapped ideal Bose gas with a few heavy impurities}

   \author{O.~Hryhorchak}
   \affiliation{Professor Ivan Vakarchuk Department for Theoretical Physics,\\ Ivan Franko National University of Lviv, 12 Drahomanov Str., Lviv, Ukraine}
   \author{V.~Pastukhov\footnote{e-mail: volodyapastukhov@gmail.com}}
    \affiliation{Professor Ivan Vakarchuk Department for Theoretical Physics,\\ Ivan Franko National University of Lviv, 12 Drahomanov Str., Lviv, Ukraine}

    \date{\today}



    \begin{abstract}
    We formulate a general scheme for calculation of thermodynamic properties of ideal Bose gas with microscopic number of static impurities immersed, when the system is loaded in the harmonic trapping potential with quasi-1D and quasi-2D configurations.
    The binding energy of a single impurity and a detailed study of the medium-induced Casimir-like forces between two impurities in trapped Bose gas are numerically calculated in wide range of temperatures and interaction strengths.
    \end{abstract}

    \maketitle

\section{Introduction}
\label{sec1}
\setcounter{equation}{0}
Recent activitization in the research field of induced forces between particles immersed in bosonic mediums is mostly stimulated by a rapid development of the Bose polaron studies in 3D \cite{Rath_2013,Zinner_2013,Grusdt_2015,Ardila_2015,Volosniev_2015,Levinsen_2015,Shchadilova_2016,Jorgensen_2016,Hu_2016,Sun_2017,Yoshida_2018,Guenther_2018,Pastukhov_2018,Mistakidis_2019,Ichmoukhamedov_2019,Yan_2020,Field_2020,Drescher_2020,Levinsen_2021,Brauneis_2021,Massignan_2021,Isaule_2021,Pascual_2021,Khan_2021,Christianen_2022,Christianen_2022_2,Schmidt_2022,Skou_2022}, in 1D \cite{Grusdt_2017,Parisi_2017,Pastukhov_2017,Kain_2018,Mistakidis_2019_2,Jager_2020,Ristivojevic_2021} and in 2D \cite{Pastukhov_2018_2,Akaturk_2019,Ardila_2020} during past decade. Particularly close to the problem of induced forces between exterior atoms in the bosonic medium is the Bose bipolaron \cite{Dehkharghani_2018,Camacho-Guardian_2018,Pasek_2019,Mistakidis_2020,Will_2021,Jager_2022,Yordanov_2023}, the problem of two (typically mutually non-interacting) impurities in the dilute Bose gas. In realistic systems, the boson-boson interaction does not allow the exact solution of the problem even in the limit of a single impurity, and only universal tail of the induced potential at large distances is accessible \cite{Schecter_2014,Naidon_2018,Camacho-Guardian_2018_2,Reichert_2019,Panochko_2022,Petkovic_2022,Fujii_2022}. The exception is mediums formed by non-interacting particles, where all details of the (in general $\mathcal{N}$-body) effective interaction can be obtained for point-like impurities. In case of free fermions, the latter leads to famous Ruderman–Kittel–Kasuya–Yosida potential. Its bosonic analogue together with the three-body inter-impurity potential were recently studied \cite{Panochko_2021} below the Bose-Einstein condensation (BEC) transition temperature in three dimensions. Thanks to simpleness of bosonic ground state in the non-interacting limit, the mean-field predictions \cite{Volosniev_2017,Panochko_2019,Hryhorchak_2020,Hryhorchak_2020_2,Guenther_2021} for energy of this system with arbitrary number of point-like impurities coincide with the exact results. The present paper generalizes these exact findings on the trapped ideal Bose gases with quasi-one-dimensional (quasi-1D) and quasi-two-dimensional (quasi-2D) geometries. There are two known facts about such a low-dimensional systems: first, the BEC exists only in the ground state; and secondly, the low-energy scattering amplitudes vanish in 1D and 2D. These all lead to a very peculiar behavior, when bosons do not experience the presence of exterior static particles at absolute zero, and only at finite temperatures the effect of impurities is visible in thermodynamics of the system.

\section{Formulation}
We consider a system of a macroscopic number $N$ of non-interacting bosons with a few $\mathcal{N}$ static (infinite-mass) impurities immersed. The initially prepared 3D system is assumed to be under the external harmonic confinement in one or two directions and for simplicity the boson-impurity interaction is taken to be the zero-range $s$-wave Huang-Yang pseudo-potential $\Phi({\bf r})=g\delta({\bf r})\frac{\partial}{\partial r}r$ (here $g=2\pi\hbar^2a/m$ with $a$ being the $s$-wave scattering length). Because there is no interaction between bosons, the ground state of system can be the non-thermodynamic one, when the immersion of microscopic number of impurities, because of formation of the boson-impurities bound states, increases energy linearly in the number of Bose particles. Before we proceed with the thermodynamic limit it is, therefore, very important to reveal all possible bound states of boson in the external `field' of impurities. 
\subsection{One-body problem}
Therefore, we consider the problem of a single trapped boson interacting with a few heavy particles. The appropriate Hamiltonian is specified as follows
\begin{eqnarray}\label{H_1}
h=-\frac{\hbar^2}{2m}\nabla^2+V({\bf r})+\sum_{1\le {\alpha}\le \mathcal{N}}\Phi({\bf r}-{\bf R}_{\alpha}),
\end{eqnarray}
where $V({\bf r})=\frac{m\omega^2}{2}z^2$ for quasi-2D, and $V({\bf r})=\frac{m\omega^2}{2}(y^2+z^2)$ for quasi-1D geometries. respectively. The set $\{{\bf R}_{\alpha}\}$ represents the positions (in general three-dimensional) of heavy impurities. An amazing fact about Hamiltonian \ref{H_1} with pseudo-potential $\Phi({\bf r})$ is that it is the exactly solvable one in general case of arbitrary number of heavy particles. Furthermore, the solution procedure can be applied for any external potential $V({\bf r})$ (for instance, the linear one \cite{Shvaika_2021}) with known eigenvalues $\varepsilon_q$ (here $q$ is the set of quantum numbers) and appropriate (normalized) wave-functions $\phi_q({\bf r})$, i.e.,
\begin{eqnarray}
-\frac{\hbar^2}{2m}\nabla^2\phi_q({\bf r})+V({\bf r})\phi_q({\bf r})=\varepsilon_q\phi_q({\bf r}),
\end{eqnarray}
(note that in those directions where the trapping potential is absent we apply the periodic boundary conditions with large length-scale $L$). Here, we are only interested in the bound states (denoting them by $\epsilon_{\mathcal{N}}$) of boson in presence of $\mathcal{N}$ interior particles, so let us define the auxiliary function 
\begin{eqnarray}\label{tilde_F}
F_{\nu}({\bf r},{\bf r}')=\sum_q\frac{\phi_q({\bf r})\phi^*_q({\bf r}')}{\varepsilon_q-\nu},
\end{eqnarray}
which is the Green function of the differential operator $-\frac{\hbar^2}{2m}\nabla^2+V({\bf r})-\nu$. Keeping the latter fact in mind as well as a singular structure of the boson-impurity interaction, the general solution of the bound-states problem can be readily constructed for an arbitrary number $\mathcal{N}$ of exterior particles
\begin{eqnarray}\label{Psi}
\Psi_{\mathcal{N}}({\bf r})=\sum_{1\le {\alpha} \le \mathcal{N}}A_{\alpha}F_{\epsilon_{\mathcal{N}}}({\bf r},{\bf R}_{\alpha}),
\end{eqnarray}
where the $\{{\bf R}_{\alpha}\}$-dependent constants $\{A_{\alpha}\}$ are subject of the boundary conditions, which formalize in the system of linear homogeneous equations
\begin{eqnarray}\label{Eq_A}
\left\{1+g\frac{\partial}{\partial r}rF_{\epsilon_{\mathcal{N}}}({\bf r}+{\bf R}_{\alpha},{\bf R}_{\alpha})\right\}_{{\bf r}=0}A_{\alpha}\nonumber\\
+\sum_{\beta\neq \alpha }F_{\epsilon_{\mathcal{N}}}({\bf R}_{\alpha},{\bf R}_{\beta})A_{\beta}=0.
\end{eqnarray}
Searching for the non-trivial solutions of this system of coupled equations we have to find zeros of its determinant. In general case these calculations can be done only numerically, but for $\mathcal{N}=1$ we write down
\begin{eqnarray}\label{Eq_1}
1+g\left\{\frac{\partial}{\partial r}rF_{\epsilon_{1}}({\bf r}+{\bf R}_1,{\bf R}_1)\right\}_{{\bf r}=0}=0,
\end{eqnarray}
while in a case of $\mathcal{N}=2$ there are two branches (note that $F_{\nu}({\bf r},{\bf r}')$ is symmetric function of its arguments for real $\nu$s)
\begin{eqnarray}\label{Eq_2}
\left[\frac{1}{g}+\left\{\frac{\partial}{\partial r}rF_{\epsilon_{2}}({\bf r}+{\bf R}_1,{\bf R}_1)\right\}_{{\bf r}=0}\right]\nonumber\\
\times\left[\frac{1}{g}+\left\{\frac{\partial}{\partial r}rF_{\epsilon_{2}}({\bf r}+{\bf R}_2,{\bf R}_2)\right\}_{{\bf r}=0}\right]\nonumber\\
-\left[F_{\epsilon_{2}}({\bf R}_1,{\bf R}_2)\right]^2=0.
\end{eqnarray}
We see that both $\epsilon_{1}$ and $\epsilon_{2}$, because of the partially broken continuous translation symmetry, depend on ${\bf R}_1$ and ${\bf R}_1, {\bf R}_2$, respectively.

\subsection{Many-body consideration}
The bosonic nature of the considered system allows for the macroscopic number of particles to be in the bound (localized) states. The total energy of the system in these collapsed BEC states is simply given by $N\epsilon_{\mathcal{N}}$.
In the following, however, we mainly focus on such a configurations of impurities, where there are no bound states in a single-boson spectrum and the ground state of the $N+\mathcal{N}$-particle system is very similar to the one without impurities. Aiming the finite-temperature description of the Bose gas with heavy particles immersed, we apply the path-integral formulation with the Euclidean action
\begin{eqnarray}\label{S}
S=\sum_{q,n} \left\{i\nu_n-\varepsilon_q+\mu \right\}\psi^*_{q,n}\psi_{q,n}\nonumber\\
-\sum_{q,q',n} g_{qq'}\psi^*_{q,n}\psi_{q',n}
\end{eqnarray}
written down in one-body basis $\phi_q({\bf r})$. In (\ref{S}) $\varepsilon_q$ denotes the sifted (on a constant term) one-particle dispersion such that $\varepsilon_{q=0}=0$ in the lowest state $q=0$; $\nu_n$ and $\mu$ stand for Matsubara frequencies ($\nu_n=2\pi n T$, where $n=0,\pm 1, \pm 2, \ldots$ and $T$ is the temperature of the system) and the bosonic chemical potential, respectively. The latter fixes density of the Bose gas. The couplings 
\begin{eqnarray}\label{g_qq}
g_{qq'}=\int d{\bf r}\phi^*_{q}({\bf r})\sum_{1\le {\alpha}\le \mathcal{N}}\Phi({\bf r}-{\bf R}_{\alpha})\phi_{q'}({\bf r}),
\end{eqnarray}
represent the matrix elements of the boson-impurities interaction. One way \cite{Nishida_2009} of dealing with (\ref{S}) is to introduce auxiliary fields that split, by means of the Hubbard-Stratonovich transformation, the last term of the above action and then integrate out the bosonic fields $\psi_{q,n}$. The remaining effective action of dummy fields is Gaussian, so the integrations can be performed to the very end. Here, however, we provide somewhat different consideration by calculating the thermal average $\langle \psi_{q,n} \psi^*_{q',n}\rangle$ explicitly. With this correlator in hand, we can obtain an equation that relates the chemical potential to the equilibrium number of bosons
\begin{eqnarray}\label{Eq_N}
N=T\sum_{q,n}e^{i\nu_n 0_{+}}\langle \psi_{q,n} \psi^*_{q,n}\rangle,
\end{eqnarray}
and taking into account `equation of motion' generated by (\ref{S}), $-\langle \psi^*_{q',n} \delta S/\delta\psi^*_{q,n} \rangle=\delta_{qq'}$:
\begin{eqnarray}\label{Eq_of_motion}
&&\left\{\varepsilon_{q}-\mu-i\nu_n\right\}\langle \psi_{q,n} \psi^*_{q',n}\rangle\nonumber\\
&&+\sum_{q''}g_{qq''}\langle\psi_{q'',n} \psi^*_{q',n}\rangle =\delta_{qq'},
\end{eqnarray}
the internal energy of the system with $\mathcal{N}$ impurities reads
\begin{eqnarray}\label{Eq_E}
E_\mathcal{N}=\mu N+T\sum_{q,n}e^{i\nu_n 0_{+}}i\nu_n\langle \psi_{q,n} \psi^*_{q,n}\rangle.
\end{eqnarray}
Being interested only in the diagonal element of correlator $\langle \psi_{q,n} \psi^*_{q',n}\rangle$, we can use adopt the Dyson-like form
\begin{eqnarray}\label{Eq_Dyson}
\langle \psi_{q,n} \psi^*_{q,n}\rangle^{-1}=\varepsilon_{q}-\mu-i\nu_n+\mathcal{T}_{qq,n},
\end{eqnarray}
found in Ref.~\cite{Luscher}. Here $\mathcal{T}_{qq,n}$ plays a role of the self energy, which is equal to the diagonal matrix element of the reduced $t$-matrix determined by following equation
\begin{eqnarray}\label{Eq_Tau}
\mathcal{T}_{qq',n}=g_{qq'}-\sum_{q''\neq q}\frac{g_{qq''}\mathcal{T}_{q''q',n}}{\varepsilon_{q''}-\mu-i\nu_n}.
\end{eqnarray}
Similarly to the translationally-invariant system of bosons and impurities (i.e., without trapping potential) \cite{Panochko_2021}, the structure of the solution can be guessed by iterating of Eq.~\ref{Eq_Tau}. Making use of the notations $\mathcal{T}_{qq',n}=\sum_{\alpha,\beta}\phi^*_{q}({\bf R}_{\alpha})T_{\alpha \beta}\phi_{q'}({\bf R}_{\beta})$, where matrix $T_{\alpha \beta}$ satisfies the linear equation
\begin{eqnarray}\label{Eq_T}
\left[\frac{1}{g}+\left\{\frac{\partial}{\partial r}rF_{\mu+i\nu_n}({\bf r}+{\bf R}_{\alpha},{\bf R}_{\alpha})\right\}_{{\bf r}=0}\right]T_{\alpha \beta}\nonumber\\
+\sum_{\gamma\neq\alpha}F_{\mu+i\nu_n}({\bf R}_{\alpha},{\bf R}_{\gamma})T_{\gamma \beta}=\delta_{\alpha \beta}.
\end{eqnarray}
This finishes a formal part of our calculations in the many-body limit at temperatures above the Bose-Einstein condensation (BEC) point. 

There is no BEC at finite temperatures in the quasi-1D or quasi-2D limits of the considered system (here we do not assume specialized thermodynamic limit, when frequency of trapping potential is sent to zero with $N\to \infty$). Note that the adopted calculation scheme is adjusted for any other external potentials, which can support the BEC transition of the system at finite temperatures. In the BEC phase, which in our case is rather interesting from methodological points of view, one has to modify the above consideration. Keeping in mind the restrictions on the configurations of impurities that do not provide the boson-impurity bound states, we have to single out the Bose condensate contributions to action (\ref{S}) $\psi_{q,n}=\sqrt{N_0/T}\delta_{q,0}\delta_{n,0}+\psi_{q,n}(1-\delta_{q,0})$ (and same for $\psi^*_{q,n}$)
\begin{eqnarray}\label{S_BEC}
S=\frac{N_0}{T}(\mu-g_{00})+\sum_{q\neq 0,n} \left\{i\nu_n-\varepsilon_q+\mu \right\}\psi^*_{q,n}\psi_{q,n}\nonumber\\
-\sqrt{\frac{N_0}{T}}\sum_{q\neq 0}\left\{g_{q0}\psi^*_{q,0} + \textrm{c.c.}\right\}-\sum_{q,q'\neq 0, n} g_{qq'}\psi^*_{q,n}\psi_{q',n}
\end{eqnarray}
where $N_0$ represents the number of bosons in BEC. Indeed, the total number of particles (\ref{Eq_N}) now implies 
\begin{eqnarray}
N=N_0+T\sum_{q\neq 0,n}e^{i\nu_n 0_{+}}\langle \psi_{q,n} \psi^*_{q,n}\rangle,
\end{eqnarray}
In the way similar to that an Eq.~(\ref{Eq_of_motion}) was obtained, we can write down for $q,q'\neq 0$
\begin{eqnarray}\label{Eq_of_motion_BEC}
&&\left\{\varepsilon_{q}-\mu-i\nu_n\right\}\langle \psi_{q,n} \psi^*_{q',n}\rangle+\sqrt{\frac{N_0}{T}}\delta_{n,0}g_{q0}\langle \psi^*_{q',0}\rangle\nonumber\\
&&+\sum_{q''\neq 0}g_{qq''}\langle\psi_{q'',n} \psi^*_{q',n}\rangle =\delta_{qq'},
\end{eqnarray}
where the nonzero average $\langle \psi^*_{q,0}\rangle$ appears due to presence of impurities. Physically, quantity $\langle \psi^*_{q,0}\rangle$ encounters for the deformation of a single-boson lowest energy wave function in the external potential of a point-like heavy particles. An average $\langle \psi^*_{q,0}\rangle$ of static part of bosonic fields can be calculated by using Eq.~(\ref{Eq_of_motion_BEC}), or equivalently derived from  $-\langle\delta S/\delta\psi_{q,0} \rangle=0$
\begin{eqnarray}
\left\{\varepsilon_{q}-\mu\right\}\langle  \psi^*_{q,0}\rangle+\sqrt{\frac{N_0}{T}}g_{0q}+\sum_{q'\neq 0}\langle\psi^*_{q',0}\rangle g_{q'q} = 0.
\end{eqnarray}
It is readily to find the solution of the above equation $\langle \psi^*_{q,0}\rangle =-\sqrt{N_0/T}\mathcal{T}_{0q,0}/(\varepsilon_{q}-\mu)$, and minimizing the grand potential with respect to $N_0$ we have
\begin{eqnarray}
\mu - g_{00} - \frac{1}{2}\sqrt{\frac{T}{N_0}}\sum_{q\neq 0}\left\{g_{q0}\langle\psi^*_{q,0}\rangle + \textrm{c.c.}\right\} = 0.
\end{eqnarray}
Combining with average $\langle \psi^*_{q,0}\rangle$, one obtains equation for chemical potential below BEC temperature
\begin{eqnarray}\label{mu}
\mu = g_{00} - \sum_{q\neq 0}\frac{\mathcal{T}_{0q,0}g_{q0}}{\varepsilon_{q}-\mu} = \mathcal{T}_{00,0}.
\end{eqnarray}
Note that similar identity $\mu=\mathcal{T}_{00,0}$ can be obtained from the condition of BEC transition as a pole of correlator (\ref{Eq_Dyson}) at zero Matsubara frequency and $q\to 0$. In general case, when the number of impurities is macroscopic Eq.~(\ref{mu}) is the transcendental one, but for our discussion ($N\gg\mathcal{N}$) the chemical potential can be freely dropped in $\mathcal{T}_{00,0}$. Remarkably, a formula for the internal energy of the system in the BEC phase is very similar to Eq.~(\ref{Eq_E}) except the zero mode in the sum over $q$ should be omitted.

\section{Results}
\begin{figure}[h!]
	\includegraphics[width=0.5\textwidth,clip,angle=-0]{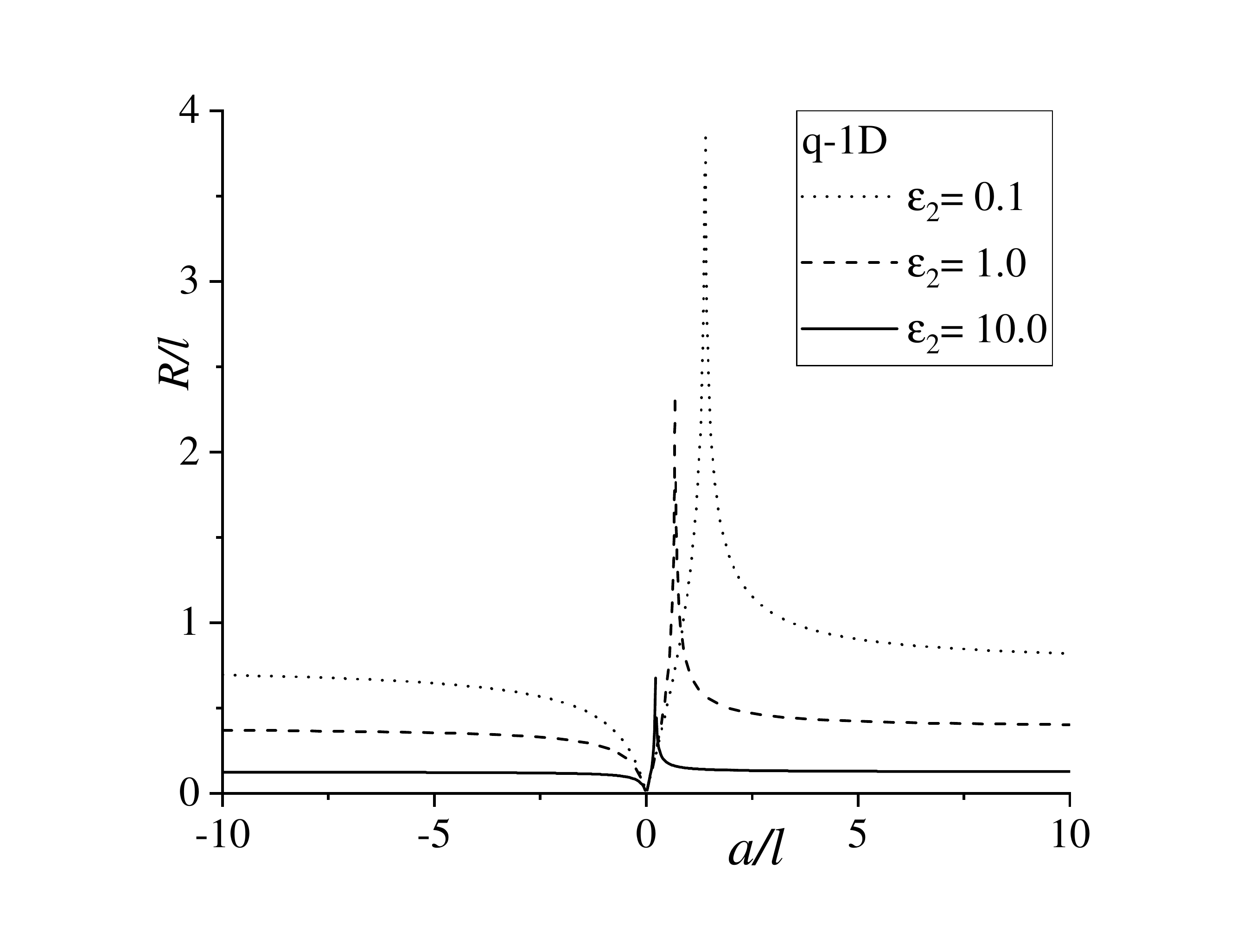}
	\caption{Bound states ($\varepsilon_2=|\epsilon_2|/\hbar\omega$) distribution of a single particle with two static impurities in quasi-1D geometry as a function of impurity separation $R$ and $s$-wave scattering length (in units of oscillator length $l=\sqrt{\hbar/m\omega}$). }\label{q1D_bound_states_fig}
\end{figure}

\begin{figure}[h!]
	\includegraphics[width=0.5\textwidth,clip,angle=-0]{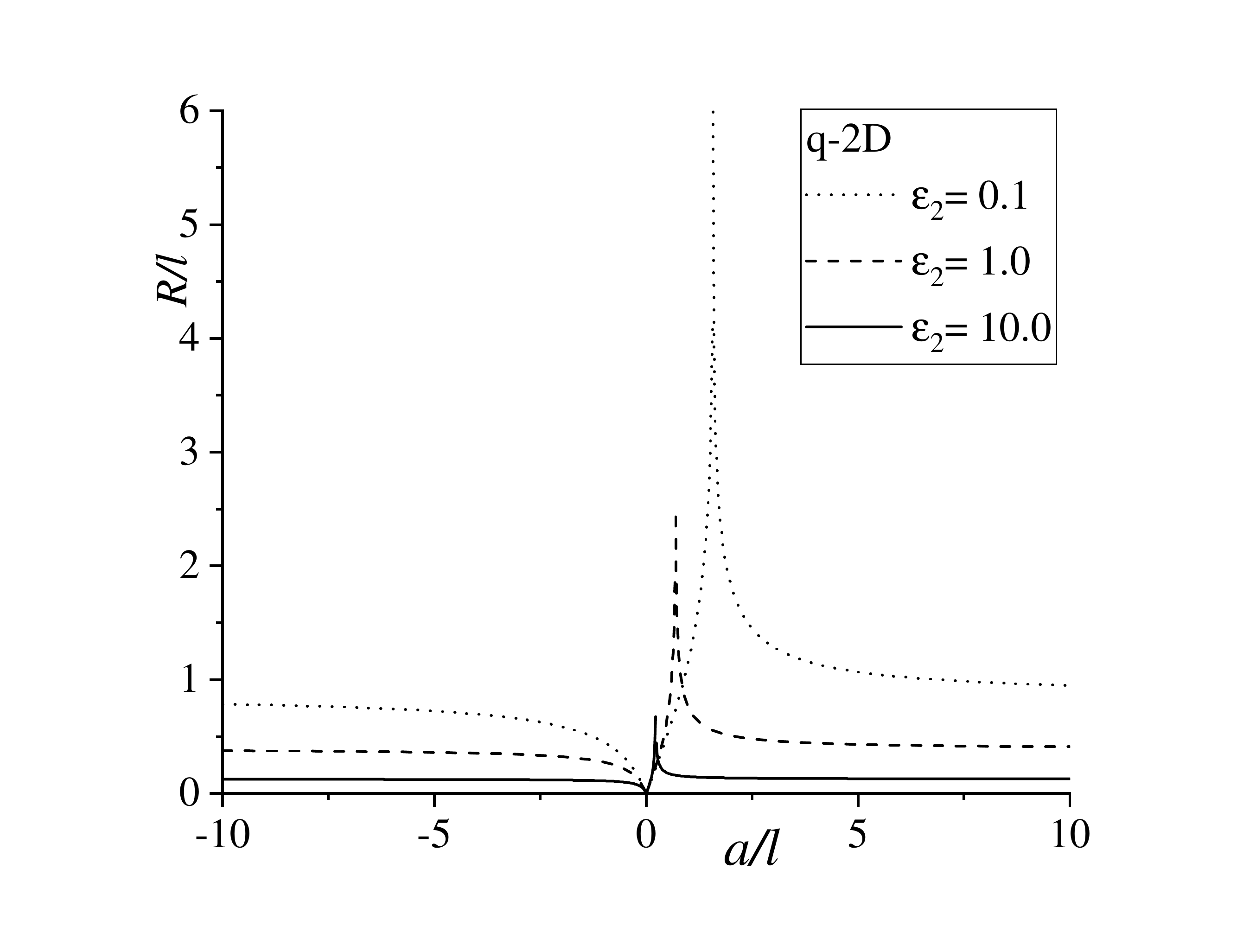}
	\caption{Same as in Fig.~\ref{q1D_bound_states_fig} but for quasi-2D case.}\label{q2D_bound_states_fig}
\end{figure}

It is naturally to start discussion of results from the bound states of a single boson in the presence of impurities. This is very important question from the point of view of thermodynamic stability of the many-body system. A detailed analysis of the two-body bound states both in quasi-1D and quasi-2D geometries were previously performed in Refs.~\cite{Olshanii,Petrov}, while here we mostly focus on a case of external potential formed by two impurities. The numerical solutions to Eq.~\ref{Eq_2} in quasi-1D and quasi-2D cases are presented in Fig.~\ref{q1D_bound_states_fig} and Fig.~\ref{q2D_bound_states_fig}, respectively. 
In order to understand the form of the bound-state energy surface we plotted the lines of equal $\epsilon_2$. Note that in both cases (quasi-1D and quasi-2D), the impurities are located at minimum of the harmonic potentials (i.e., on the $x$-axis in quasi-1D geometry and in the $xy$-plane in quasi-2D case), because otherwise the bound states and thermodynamic characteristics are exponentially suppressed. It is readily seen from Figs.~\ref{q1D_bound_states_fig}, \ref{q2D_bound_states_fig} that the qualitative picture is the same for two geometries: there are two branches (by number of impurities) of bound states for positive $a$s and one for negative. If we increase number of impurities immersed to $\mathcal{N}$, there will be exactly $\mathcal{N}$ bound states for $a>0$ and maximally $\mathcal{N}-1$ bound state for negative $a$s. This distribution of branches is completely analogous \cite{Panochko_2021} to the translation invariant three-dimensional case.

\begin{figure}[h!]
	\includegraphics[width=0.5\textwidth,clip,angle=-0]{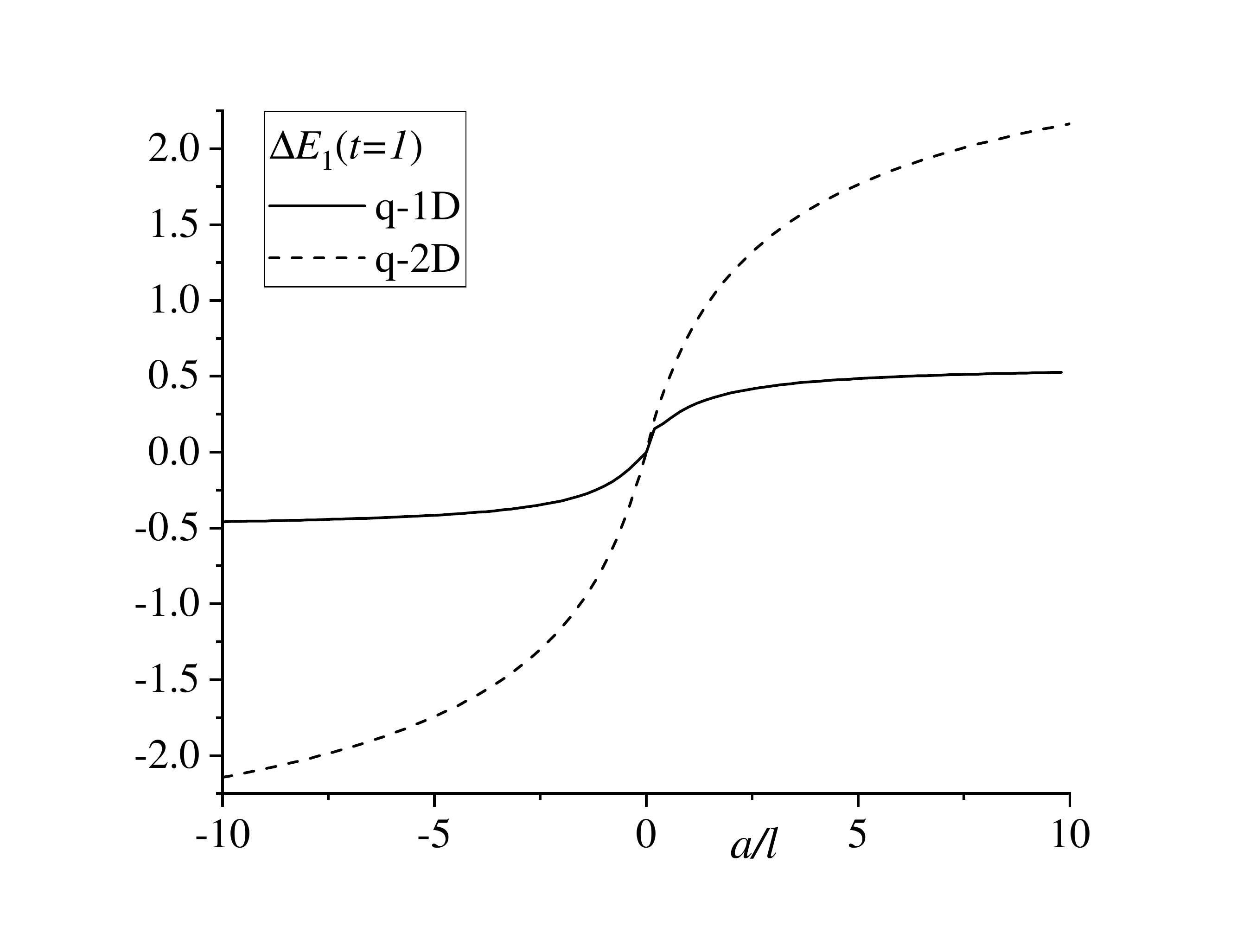}
	\caption{One-impurity contribution (in units of $\hbar\omega$) to the internal energy of quasi-1D and quasi-2D Bose gases at finite temperature $t=T/\hbar\omega$.}\label{DeltaE_1_fig}
\end{figure}

\begin{figure}[h!]
	\includegraphics[width=0.5\textwidth,clip,angle=-0]{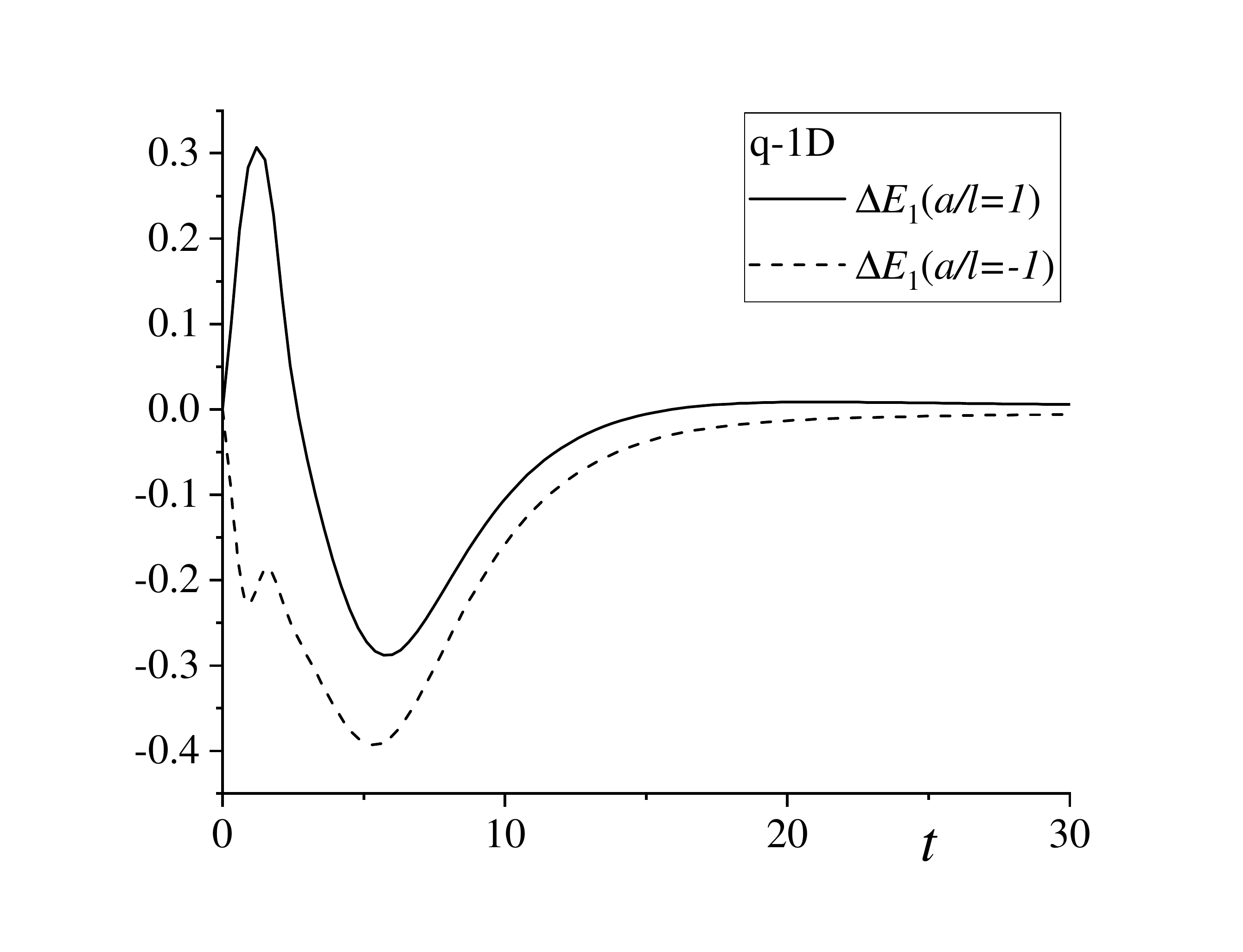}
	\caption{Temperature dependence of the one-impurity impact (in units of $\hbar\omega$) to the internal energy of quasi-1D ideal Bose gas.}\label{DeltaE_1_q1D_fig}
\end{figure}

\begin{figure}[h!]
	\includegraphics[width=0.5\textwidth,clip,angle=-0]{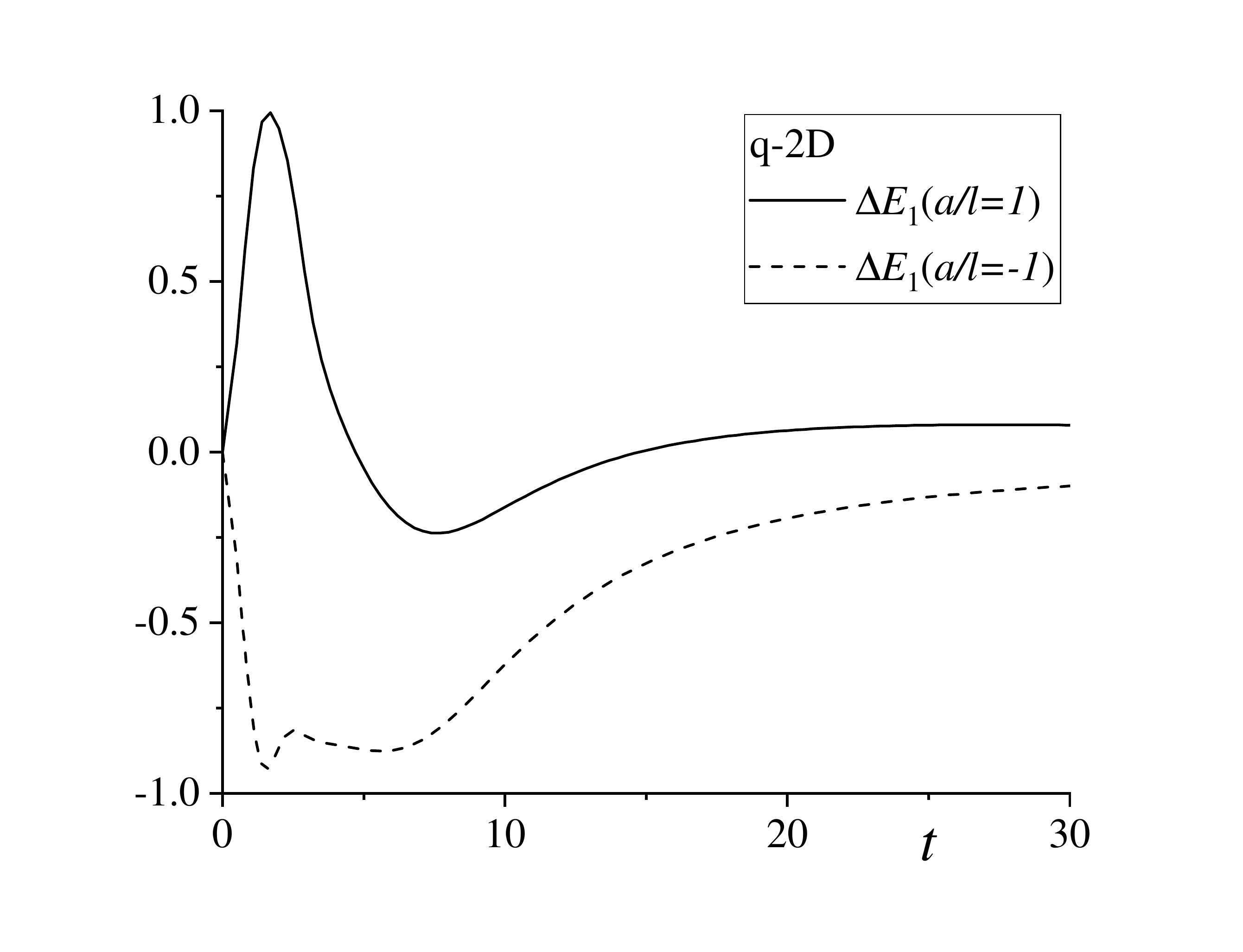}
	\caption{Same as in Fig.~\ref{DeltaE_1_q1D_fig}, but for quasi-2D geometry.}\label{DeltaE_1_q2D_fig}
\end{figure}

\begin{figure}[h!]
	\includegraphics[width=0.5\textwidth,clip,angle=-0]{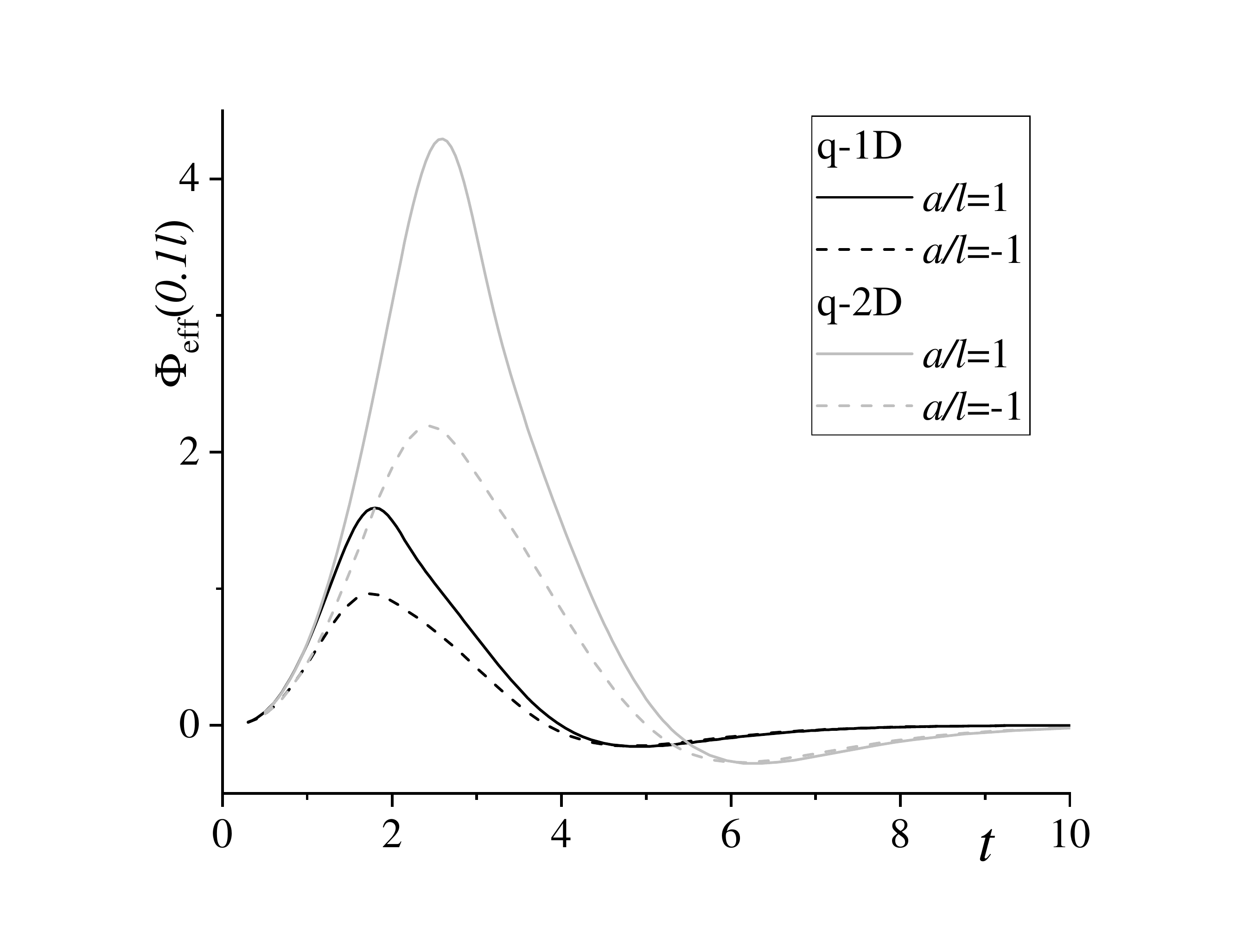}
	\caption{Temperature dependence of the effective impurity-impurity potential at $R=0.1l$.}\label{Phi_t_fig}
\end{figure}

Talking about thermodynamic properties of our system it should be clearly understood that in their non-BEC-collapsed ground state (which is not the thermodynamic one) free bosons in low dimensions are insensible to the presence of static impurities. This happens because the 1D and 2D $t$-matrices vanish at zero energy of colliding particles. And only thermally-stimulated bosons that scatter on impurities impact to the energy of the system at finite temperatures. Another way to provide a non-zero population of excited states (and consequently, non-zero immersion energy of impurities) is to turn on the interaction between bosons \cite{Panochko_2022}. In Fig.~\ref{DeltaE_1_fig}
we plotted a typical dependence of correction to the internal energy on $s$-wave scattering length (in units of oscillator length $l=\sqrt{\hbar/m\omega}$) of ideal Bose gas caused by a single impurity in quasi-1D and quasi-2D geometries at temperature $t=T/\hbar \omega=1$ (for our numerical computations we fixed the densities of quasi-1D and quasi-2D Bose gases as $n_{1D}l=1$ and $n_{2D}l^2=1$, respectively). Although, the general behavior of curves is qualitatively the same, a spatial dimensionality impacts crucially to the correction magnitude. We have also analyzed the temperature dependence of the one-impurity correction to the thermodynamics of ideal Bose gas at fixed scattering length ($a=\pm l$) in quasi-1D \ref{DeltaE_1_q1D_fig} and quasi-2D \ref{DeltaE_1_q2D_fig} cases.
An important conclusions of these calculations are that the one-impurity correction depends non-monotonically on the temperature and character of curves is different for attractive ($a<0$) and repulsive ($a<0$) boson-impurity interactions. We think the former results is due to presence of trapping potential, because such a non-monotonicity was not observed \cite{Panochko_2021} in truly 3D case at least in BEC region. For the determination of the effective induced two-body potential between infinitely heavy particles we calculate the internal energy Eq.~(\ref{Eq_E}) of bosonic system with exactly two impurities (put at a distance $R$ one from another) immersed and substitute the doubled correction to internal energy of a system caused by a single impurity (in other words, substitute two-impurity energy with $R\to \infty$) 
\begin{eqnarray}
\Phi_{\textrm{eff}}(R)=E_2(R)-E_2(\infty).
\end{eqnarray}
\begin{figure}[h!]
	\includegraphics[width=0.5\textwidth,clip,angle=-0]{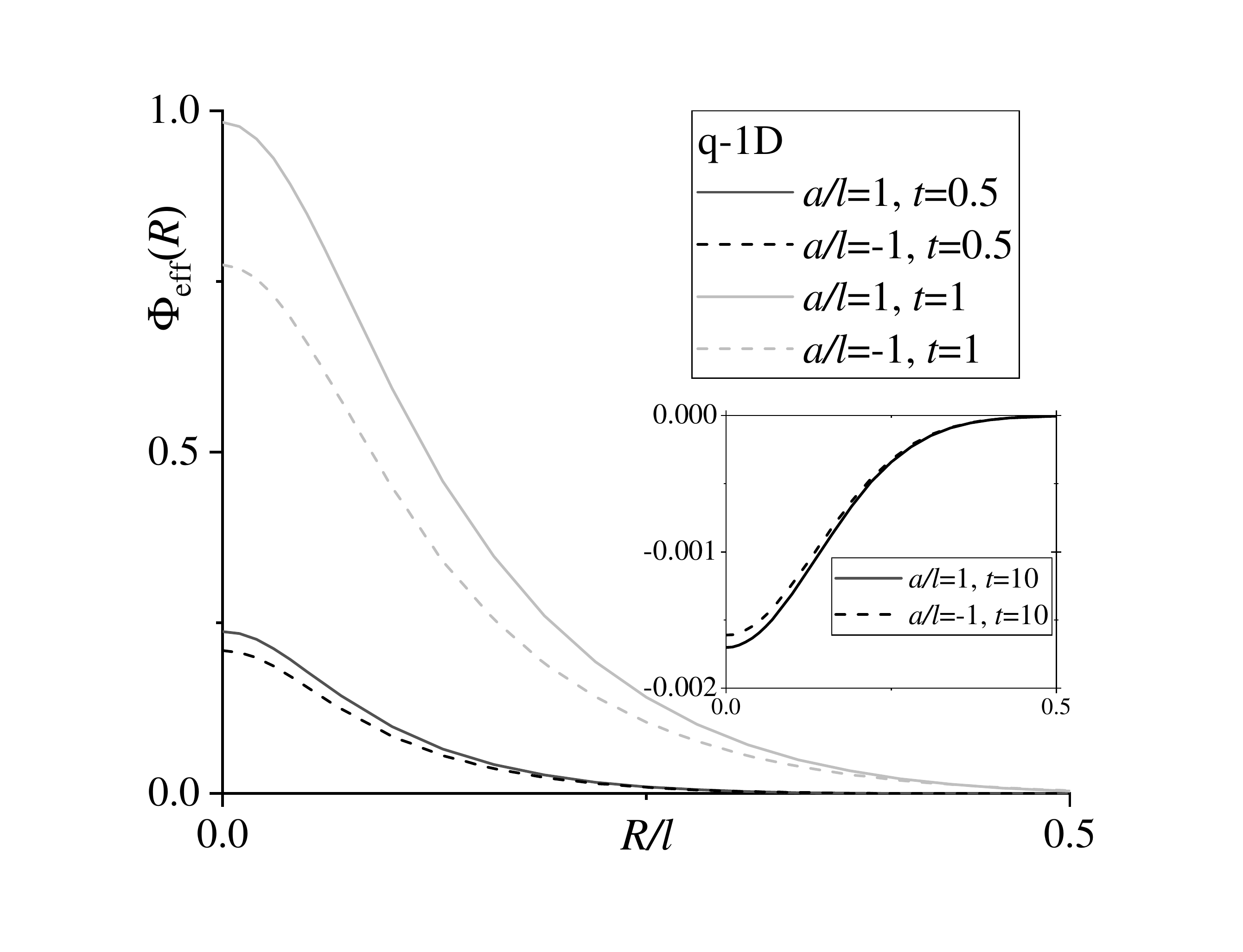}
	\caption{The effective medium-induced two-body potential (in units of $\hbar\omega$) for static impurities immersed in quasi-1D ideal Bose gas.}\label{Phi_R_1D_fig}
\end{figure}
\begin{figure}[h!]
	\includegraphics[width=0.5\textwidth,clip,angle=-0]{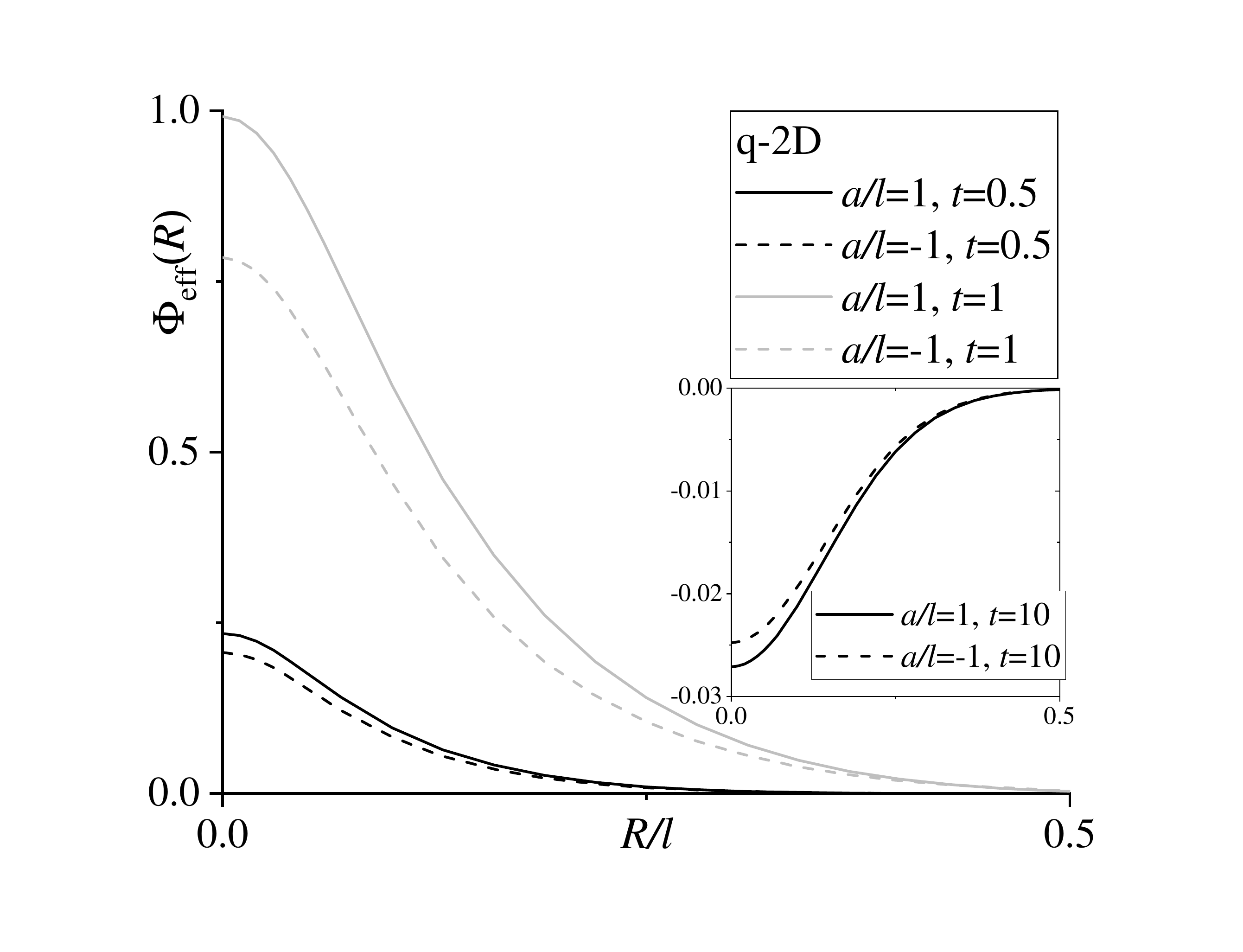}
	\caption{Same as in Fig.~\ref{Phi_R_1D_fig}, but for quasi-2D geometry.}\label{Phi_R_2D_fig}
\end{figure}
The results of numerical calculations of $\Phi_{\textrm{eff}}(R)$ for quasi-1D and quasi-2D cases are presented in Fig.~\ref{Phi_R_1D_fig} and Fig.~\ref{Phi_R_2D_fig}, respectively. Let us make two comments about obtained behavior of $\Phi_{\textrm{eff}}(R)$. First, it seen that in both quasi-1D and quasi-2D cases, the character of curves is very similar even quantitatively for same sets of parameters, and visible discrepancies appear only in the high-temperature region (compare insets in Figs.~\ref{Phi_R_1D_fig},\ref{Phi_R_2D_fig}). Secondly, as a function of temperature the effective two-body potential changes its sign from the repulsive at low temperatures to the attractive in the high-temperature region. In order to visualize this pattern, we plotted $\Phi_{\textrm{eff}}(R=0.1l)$ in Fig.~\ref{Phi_t_fig}
at various temperatures. The obtained curves particularly suggest the emergence of the thermally-stimulated two-body bound state of impurities with finite (but large) masses, but to find the solution to this problem one needs to go beyond the approximation of static impurities.

\section{Summary}
In conclusion, we have presented a detailed analysis of the impact of one and two static impurities on properties of the harmonically trapped ideal quasi-1D and quasi-2D Bose gases. Within the assumption about a short-range character of the boson-impurity interaction, the formulated general scheme allows the calculation of thermodynamics of the system in any external potential and with the arbitrary impurity number. Particularly, we have identified the detailed dependence of the energy of a single impurity immersed in the quasi-low-dimensional trapped Bose gas on temperature and interaction strength. The calculations of the medium-induced effective potential between two impurities in the system of free bosons revealed an interesting behavior: repulsion at low temperatures that changes to attraction in the high-temperature limit. As a byproduct of this study, we have identified the bound states of a single particle interacting through the Huang-Yang pseudopotential with two static impurities separated at arbitrary distances in quasi-1D and quasi-2D geometries.

\section*{Acknowledgements}
This work was partly supported by Project No.~0122U001514 from the Ministry of Education and Science of Ukraine.

\end{document}